\documentclass[aps,prd,twocolumn,superscriptaddress,
amsmath,amssymb,nofootinbib
]{revtex4-2}

\usepackage{slashed}
\usepackage{color}
\usepackage[breaklinks,colorlinks=true]{hyperref}

\usepackage{braket}
\usepackage{mathrsfs}

\newcommand{\dd}[0]{\partial}
\newcommand{\tr}[0]{\text{tr}}

\newcommand{\bea}{\begin{eqnarray}}
\newcommand{\eea}{\end{eqnarray}}
\newcommand{\beq}{\begin{equation}}
\newcommand{\eeq}{\end{equation}}

\begin{document}
\title{Qubit Casimir effect}

\author{Antonino Flachi}
\affiliation{Department of Physics \& Research and Education Center for Natural Sciences,
  Keio University, 4-1-1 Hiyoshi, Yokohama, Kanagawa 223-8521, Japan}

\author{Gon\c{c}alo M. Quinta}
\affiliation{Physics of Information and Quantum Technologies, Instituto de Telecomunica\c{c}\~{o}es, Lisboa, Portugal}

\begin{abstract}
In this letter we propose a new interpretation of the Casimir effect. Concretely, we show that the Casimir energy can be written as the quantum ``Von Neumann'' entropy associated to a 2-qubit, mixed pseudo-density matrix of the relevant quantum fluctuations. The quantum entropy we introduce draws parallels to the concept of quantum inseparability found in quantum information theory. Our results suggest that the Casimir energy is a measure of the quantum entropy of the vacuum fluctuations.
\end{abstract}
\maketitle

\textit{Introduction} -- Intrigued by a suggestion due to Bohr, Casimir showed that the retarded dispersion force between two polarizable molecules, a result obtained earlier with Polder \cite{CasimirPolder:1948}, could be derived by comparing the zero-point energy for the electromagnetic field in the presence of the molecules and in vacuum. This analysis, when applied to parallel flat boundaries, yields Casimir’s renowned result on the attractive forces between two perfectly conducting plates, known as the Casimir effect \cite{Casimir:1948dh}. This was a pivotal conclusion of quantum electrodynamics and has sparked considerable debate about the significance 
of zero-point fluctuations \cite{Milonni:2000yj,Jaffe:2005xx}.
Experimental measurements of the Casimir effect have largely mitigated skepticism regarding this phenomenon and supported the notion of real zero-point fluctuations, despite Schwinger’s alternative derivation through source theory, which does not invoke zero-point energies \cite{Schwinger:1975ceff}. Moreover, the Casimir effect can be entirely articulated through S-matrix elements, sidestepping zero-point energy \cite{Bordag:1983zk,Peterson:1981yy,Jaffe:2005xx}. An $S$-matrix derivation does necessarily include virtual quanta across propagators and loops, and once we connect the Casimir energy to quantum source fluctuations, ``virtuality'' seems unavoidable in Schwinger's approach as well \cite{Milonni:2000yj}. 

In tabletop experiments, interpreting zero-point energies often takes a backseat to matching theory with measurements. {(The first successful experiment with parallel plates involved opposing surfaces of a cantilever beam, which could oscillate freely around its fixed point, and a thicker beam firmly attached to a frame, allowing for an adjustable distance from the cantilever \cite{Bressi:2002fr}. Earlier successful experiments utilized a sphere and plate configuration, with torsion balances \cite{Lamoreaux:1996wh}, atomic force microscopes \cite{Mohideen:1998iz}, and highly precise capacitance bridges \cite{Chan:2001zzb} as measuring devices. See Ref.~\cite{Bordag:2009zz} for further details and references and Refs.~\cite{Norte:2018sc,Xu:2022sp,Xu:2022nt} for a few recent unconventional examples.)} However, in the absence of experimental data, the nature and significance of zero-point fluctuations remain elusive, especially regarding their gravitational effects and the puzzling cosmological constant. That is, once we acknowledge that a finite Casimir energy gravitates as required by the equivalence principle \cite{Fulling:2007xa}, Lorentz invariance implies an energy-momentum tensor proportional to the energy density of the vacuum, $\langle \rho \rangle $. 
Since $\langle \rho \rangle \sim \left[\mbox{Planck scale}\right]^4$, 
we are urged to explain why the observed value of the cosmological constant is much smaller. 
Although unsuccessful, Zeldovich was the first to think about the possibility of a cancellation mechanism of this large vacuum energy, leaving higher order contributions and a small cosmological constant, anti\-cipating the more evident fine tuning issue  
occurring in electroweak theory \cite{Zeldovich:1967cp,Weinberg:1988cp}.

Defining a renormalized Casimir energy as the difference between zero-point energies with and without boundaries aligns with normal ordering and yields a finite result. However, it falls short in the cosmological constant problem, resulting in vanishing vacuum energy in the thermodynamic limit. A more intuitive picture, grounded in the Heisenberg uncertainty principle, suggests that the Casimir force arises from an imbalance between inward and outward pressures on the boundaries caused by these fluctuations, as a result of {the decay of the quantum vacuum state}. 
All this suggests that our current interpretation might be missing a crucial piece of the puzzle, and these questions inspired us to explore a different perspective on the Casimir effect, leading us to interpret it in terms of quantum entropy.

\textit{Casimir energy as a functional determinant} -- 
Let us consider a $d=(1+p)$-dimensional, ultra-static Euclidean background $\Sigma$ with a metric $ds^2 = d\tau^2 +g_{ij}dx^i dx^j$; $\tau$ is the Euclidean time $t\to {-}i\tau$, with $\tau \in \mathbb{S}^1(\beta)$ of radius $\beta$, which is let to $\infty$ (zero temperature limit) at the end of the calculations. We assume $\Sigma$ to be a compact manifold of volume $V$ with codimension-1 boundaries $\partial \Sigma$. (For example, $\Sigma$ could be flat space and $\partial \Sigma$ a set of two parallel disconnected flat boundaries plus periodic boundary conditions imposed in the unconfined directions.) Let us consider a differential operator of the form 
\bea
\mathscr{D}_\Phi=\left(\partial_\tau^2 + D\right) \delta^{i}_j   + \delta^{i}_k \Phi^k_{j}(x),
\label{DE}
\eea 
with $D$ being a second order, self-adjoint, elliptic differential operator, $\delta^{i}_k$ the $N$-dimensional identity, $N$ a positive integer and $\Phi^k_{j}(x)$ a matrix valued field. The latin indices run from $1$ to $N$, over the field multiplicity. 
Finally, let us define the eigen-frequencies associated with the spatial part of the operator, $\mathsf{D}_\Phi= D\;\delta^{i}_j+\Phi^k_{j}\delta^{i}_k$, to be $\epsilon = \sqrt{\lambda}$. For notational convenience we have dropped the indices from the eigenvalues and from the matrix differential operators and identified $\mathscr{D}_\Phi \equiv \left(\mathscr{D}_\Phi\right)^i_j$ and $\mathsf{D}_\Phi \equiv \left(\mathsf{D}_\Phi\right)^i_j$.

Our focus 
is the Casimir energy $E_c$ (we set $\hbar=c=1$) 
\bea
E_{c} = 
{1 \over 2} \sum_{\epsilon} \epsilon,
\eea
where the above divergent sum runs over the whole spectrum and zero eigenvalues are not included. Utilizing zeta regularization, we define a regularized Casimir energy by analytic continuation to $s\to 0$ of
\bea
E_{c} &=&  
\lim_{s\to 0}  
{1 \over 2} \sum_{\epsilon} \epsilon \left( \ell\epsilon\right)^{- 2s}
= \lim_{s\to 0}  
{\ell^{-2s} \over 2} \zeta\left[{\mathsf{D}_\Phi}\right]\left(-{1\over 2}+ s\right);~~~
\label{zetaE}
\eea
$\ell$ is a regulator with dimensions of length and $\zeta\left[{\mathsf{D}_\Phi}\right](z) = \sum_{\lambda} \lambda^{-z}$ the zeta function associated with $\mathsf{D}_\Phi$. 
The zeta function (\ref{zetaE}) is meromorphic with at most simple poles that can be extracted using the Mittag-Leffler representation of a second order operator ${\mathsf{P}}$ \cite{Gilkey:1984it,sergei:1994,Kirsten:2001wz,Blau:1988kv},
\bea
\zeta\left[{\mathsf{P}}\right]\left(s\right) = {1\over \Gamma(s)} \left\{\sum_p {C_{p}\left({\mathsf{P}}\right)\over s - p +D/2} + {f}(s)
\right\},
\label{mitlef}
\eea 
where $C_{p}$ is the $p$-th heat-kernel coefficient, written in terms of integrals of geometrical invariants $C_{p} = \int_{\Sigma} a_{p} + \int_{\partial \Sigma} b_{p}$ with $a_{p}$ and $b_{p}$ being the volume and boundary heat-kernel coefficients, respectively \cite{Gilkey:1984it,Kirsten:2001wz}; $f(s$) is an entire function. In general, the presence of such poles in the zeta function does not allow to remove the regulator. In flat space and for flat boundaries, $C_{d/2} =0$ and the analytic continuation to $s\to 0$ yields a finite result independent of the renormalization scale. When $C_{d/2} \neq 0$ renormalization of the bare action is required. 
An unambiguous definition can always be given for a renormalizable quantum field theory, e.g., the minimal subtraction scheme corresponds to taking the principal part of (\ref{zetaE}) that removes the pole part leaving a finite result.

The Casimir energy can also be obtained using the effective action. For the operator $\mathscr{D}_\Phi$ the one-loop effective action is \cite{Kirsten:2001wz,Parker:2009uva,Blau:1988kv,Fosco:2008vn}
\bea
\Gamma^{(1)} = {1\over 2} \ln \det \mathscr{D}_\Phi = - {1\over 2} \zeta\left[{\mathscr{D}_\Phi}\right]'(0)
\label{seff}
\eea
where the zeta function can be written in terms of the heat-kernel using the Mellin transform (as before, zero eigenvalues are excluded):
\bea
\zeta\left[{\mathscr{D}_\Phi}\right](s) &=& 
{1\over \Gamma(s)} \int_0^\infty {dt\over t^{{1-s}}} \sum_{\xi} e^{-t \ell^{-2}\xi},~~~~
\eea
with $\xi=\omega_n^2 + \epsilon$, where $\omega_n$ are the eigenvalues of $\partial_\tau^2$. Since we are dealing with a second order scalar operator we impose periodic boundary conditions along $\tau$. Here, we are not interested in finite temperature corrections and will take the zero temperature limit. The integrand is the diagonal part of the heat-kernel, $\tr\left(\exp (-t \mathscr{D}_\Phi \ell^2)\right) = \sum_{\xi} \exp(-t \ell^{-2}\xi)$ \cite{Kirsten:2001wz}.
For an ultrastatic spacetime and the operator in (\ref{DE}), the heat-kernel factorizes allowing us to relate $\zeta\left[{\mathscr{D}_\Phi}\right](s) $ with $\zeta\left[{\mathsf{D}_\Phi}\right](s)$
\cite{Toms:2007sch,Blau:1988kv}:
\bea
\zeta\left[{\mathscr{D}_\Phi}\right](s)  = {1\over \sqrt{4\pi} \ell}
{\Gamma(s-1/2)\over \Gamma(s)}
\zeta\left[{\mathsf{D}_\Phi}\right](s-1/2).
\label{zeta34}
\eea
Using (\ref{seff}) and (\ref{zeta34}) we can obtain the one-loop effective potential, $\mathsf{E_{eff}} = -\beta^{-1} V^{-1} \Gamma^{(1)}$ in $d=4$ ($V$ is the volume). For a Casimir system, one is usually interested in the energy per unit boundary-area $S$, thus defining $\mathsf{E_{eff}} = -\beta^{-1} S^{-1} \Gamma^{(1)}$. Using (\ref{zeta34}), (\ref{seff}) and (\ref{mitlef}) one has
\bea
\mathsf{E_{eff}} = E_c +{C_2 \over 32 \pi^2 \ell} \left(\psi(1)-\psi(-1/2)\right)
\label{eeff}
\eea
where $\psi(s)$ is the digamma function. When $C_2=0$, there is no ambiguity: 
$\mathsf{E_{eff}}$ and $E_c$ coincide to a finite (zeta-regularized) value with no renormalization of diverging contributions being required. 
If the background or the boundaries are curved the ambiguity needs to be removed by a choice of a renormalization conditions. 
In the following we will focus on the setup involving two ideal parallel plates of area $S$ at a distance $L$, in which case the result is finite (see equations (\ref{mitlef}) and (\ref{eeff})). Furthermore, it is supported by experimental measurements conducted within the context of the electromagnetic field \cite{Bressi:2002fr}. For this case, the calculation of the zeta function is a textbook result (See Ref.~\cite{Toms:2007sch,sergei:1994,Blau:1988kv,Milton:2001xx}) leading to 
the following Casimir energy per unit area $S$ per degree of freedom $N$
\bea
\mathscr{E}_c = E_c / (NS) = - \pi^2/ (1440 L^{3}).
\label{parplates}
\eea

\textit{Replica trick} -- A qubit framework is intuitive for electrons or photons, where two levels can be made to correspond to helicity or polarization; however, it is not universally applicable. To introduce a qubit description of the Casimir effect, we must first establish a method to associate a discrete ``two-state structure'' with the virtual fluctuations, which may not have an explicit spin structure.

We can assign qubits to a functional determinant using a replica trick, i.e., by expressesing the determinant (\ref{seff}) in terms of the eigenvalues of a Dirac operator. This can be done using the Schrödinger-Lichnerowicz-Peres relation \cite{Parker:2009uva,Gibbons:2019coc},
\bea
\det \left(\ell^2 \left(\Box +m^2 +{1\over 4}R \right)\right)
=
\det{}^2 \left(\ell (i \underline{\gamma}^\mu \nabla_\mu -m) \right),~~~~~~
\label{SLPrel}
\eea
where the operator on the left is a diagonal matrix operator of dimensionality ${\cal{N}}$ with ${\cal{N}}$ equal to the dimensionality of the relevant Clifford algebra. (In general, one can always attach a qubit structure to a determinant independently of the native spin structure by considering $n$ copies of the original theory with $n$ chosen appropriately.)
The gamma matrices $\underline{\gamma}^\mu$ and the covariant derivative are defined in curved space through vierbeins and the connection \cite{Parker:2009uva}. 
Here, we shall focus on $d=(1+3)$ and identify $\Phi^k_j \equiv (m^2 + R/4) \delta^k_j \equiv \Phi \delta^k_j$. Then, the determinant in (\ref{seff}) can be written in terms of that of the Dirac operator as in (\ref{SLPrel}). Taking the logarithm of the RHS of (\ref{SLPrel}) gives, in the limit of $\beta \to \infty$, the zero temperature one-loop effective action: 
\bea
\Gamma^{(1)} = 2 \beta V \sum_{\sigma} \sqrt{\sigma + \Phi},
\label{aven} 
\eea 
where $\sigma = \lambda - \Phi$ are the eigenvalues of $D$ (see (\ref{DE})). If we set the background to be flat and $m\to 0$ (i.e., $\Phi\to 0$), we have the usual massless Casimir energy expression (multiplied by the volume) $\sim \sum_{n} \sqrt{\mathbf{k}_n^2}$, i.e. proportional to the sum over the quantized momenta $\mathbf{k}_n$.
The quantity (\ref{aven}) defines the average energy at zero temperature, as it can be seen directly from the partition function $Z$,
as {$\braket{\mathsf{E}} = -{\partial \over \partial \beta} \ln Z = - 2 \sum_{\sigma} \sqrt{\sigma + \Phi}$}. The formula above reiterates that at zero temperature, the free energy is proportional to the vacuum energy (see also relation (\ref{eeff})).
We stress here that an ordinary definition of (thermodynamic) entropy, i.e. $S_T = \beta^{-2} {{\dd} F}/{{\dd} \beta}$, with  {$F = - \beta^{-1} \ln Z(\beta)$} being the free energy, approaches zero in the limit of zero temperature $\beta \to \infty$, which aligns with the Nernst theorem applicable to regular bodies. Thus, $F = \braket{\mathsf{E}} - T S_T$ implies that at zero temperature $T$, the only accessible energy is the Casimir energy.

\textit{Pseudo-density matrix} -- In the following 
we shall limit ourselves to the parallel plates' setup 
discussed earlier and to fermion fluctuations with the \textit{proviso} that, at least in absence of self-interactions, formula (\ref{SLPrel}) allows one to obtain the Casimir energy for other spins, modulo a numerical multiplicative factor. Taking the fermion determinant as a starting point and moving to momentum space, we have 
\cite{Kapusta:2006pm}
\bea
\Gamma^{(1)} = \ln \det \check{\rho}_k
\label{fdet}
\eea
where we have defined
\bea
\check{\rho}_k = {\left( \slashed{k}+m  \right)\gamma^0\over 4 k^0} \equiv {1\over 4}\left(I + \chi_{k} B\right).
\label{pseudorho}
\eea
The functional trace in (\ref{fdet}) ($\ln \det = \tr \ln$) is carried over all degrees of freedom compatibly with the boundary conditions, i.e. summations implicit in the trace are done over the appropriately quantized momenta (with zero eigenvalues removed). The last equality sign in (\ref{pseudorho}) defines the matrix $B$, 
\bea
B = {m - k_i \gamma^i \over k^0 \chi_{k}}\gamma^0.
\label{pseudoB}
\eea
with 
\bea
\chi_{k} = \omega_{\bf k} /k^0 = \sqrt{{\bf k}^2 + m^2}/ (i \omega_n).
\label{chik}
\eea 
The matrix $B$ is involutory, $B^2=I$ and traceless (as it follows from the properties $\tr \gamma_{\mu}= \tr \gamma_i\gamma^0=0$). We follow the usual notation $\slashed{k} = k^\mu \gamma_\mu = k_0 \gamma^0 - k_i \gamma^i$ and  define {$k^\mu = \left(i \omega_n, {\mathbf k}_{\mathbf j} \right)$}, where ${\mathbf j}$ is a spatial multi-index fixed according to the boundary conditions. In $3$ spatial dimensions and for the parallel plates problem with Dirichlet boundary conditions imposed at $z=0$ and $z=L$ and periodic boundary conditions in the other directions, we have ${\mathbf k}_{\mathbf j} = \left( 2 \pi j / L_x ,  2 \pi k / L_y,  \pi l / L_z \right)$ with {$j,k \in \mathbf{Z}$ and $0< l \in \mathbf{N}$}. 
Notice that all we have done in (\ref{fdet})-(\ref{pseudorho}) has been re-writing the usual expression 
by factorizing out the term $k^0$; this has introduced the factor $1/k^0$ (this manipulation is possible because a multiplicative $k^0$ factor in the determinant does not contribute to the Casimir force, as it does not depend on $L$ and thus disappear when taking the derivative with respect to $L$). The coefficient $1/4$ is introduced for convenience. The advantage of expressing the determinant as in (\ref{fdet}) - (\ref{pseudorho}) is that $\check{\rho}_k$ can be written 
in thermal form (see \cite{Bjorken:1964rqm,Quinta:2023qb,Quinta:2023pkx} for a general proof that we revisit here). Starting from $\left(I + \chi_{k} B\right)/4 = A_k \exp\left(\alpha_k M \right) = A_k \left(I \cosh \alpha_k + M \sinh\alpha_k\right)$ where $A_k$ and $\alpha_k$ are scalars and $M$ is involutory, one can derive the following relations: $M=B$, $A_k \cosh \alpha_k =1/4$ and $A_k \sinh \alpha_k =\chi_k/4$; from these it is easy to arrive at $A_k = \sqrt{1-\chi_k^2}/2$ and $\alpha_k = \mbox{artanh} \chi_k = \log\left({(1+\chi_k)/(1-\chi_k)}\right)/2$. Finally, utilizing $A_k  \tr \exp\left(\alpha_k M \right) = A_k \times 4 \cosh \alpha_k = 1$, we arrive at the following expression:
\bea
\check{\rho}_k = {\exp \left(- \check{\beta}_k \mathscr{H}_k \right)\over \tr \exp\left(- \check{\beta}_k \mathscr{H}_k \right)}
\label{pseudorhodens}
\eea
where
\bea
\check{\beta}_k &=& {1\over 2 k^0}\log\left({(1+\chi_k)/(1-\chi_k)}\right) \label{pseudobeta}
\eea
is a \textit{real} scalar quantity with units of $\left[\mbox{energy}\right]^{-1}$ that we identify with the inverse of a pseudo-temperature (notice that this pseudo-temperature becomes negative for $\omega_{\bf k}/\omega_n < 0$ and vanishes for $\omega_{\bf k}/\omega_n = 0$) and
\bea
\mathscr{H}_k &=&\left({k_i \gamma^i - m}\gamma^0\right)/\chi_k. \label{pseudoH}
\eea
is the fermion Hamiltonian modulated by the dimensionless coefficient $\chi_k$. Relation (\ref{pseudorhodens}) is formally valid for any complex $k$. The matrix $\check{\rho}_k$ can be written as sum of an hermitian plus an anti-hermitian parts; its eigenvalues are $\lambda^{(\pm)}_{k} = (1\pm \chi_{k})/4 \in \mathbb{C}$, each with multiplicity $2$, as it can be shown by directly computing the spectrum of (\ref{pseudorho}). Written as (\ref{pseudorhodens}), $\check{\rho}_k$ represents the contribution from each $k$-mode, each characterised by a set of quantum numbers labeling the quantized momenta ${\mathbf k}$ and frequency $\omega_n$. 
In fact, $\check{\rho}_k$ only satisfies normalization conditions, but it is not hermitian nor semi-positive definite, as one can immediately observe from its eigenvalues and from (\ref{chik}). In fact, it is not difficult to see that $\check{\rho}_k$ is pseudo-hermitian, i.e. hermitian with respect to a nontrivial metric (it satisfies the relation $\mathbb{B} \check{\rho}_k^\dagger =  \check{\rho}_k \mathbb{B}$ with $\mathbb{B} = \gamma^1\gamma^2\gamma^3$. See, for example, Ref.~\cite{Mostafazadeh:2008pw} for details on pseudo-hermitian matrices). The unusual properties of $\check{\rho}_k$ are expected and due to the fact that the pseudo-density matrix is associated to virtual degrees of freedom that are not on-shell, i.e., fluctuations which may travel back and forth in time. To highlight the difference with ordinary density matrices, we introduced the term \textit{pseudo-density} matrix for $\check{\rho}_k$.

We can achieve a better understanding of the pseudo-density matrix by looking at its eigen-vectors.
An elementary calculation shows that the eigen-vectors of $\check{\rho}_k$ can be obtained by boosting 
in the $+{\bf k}$ ($-{\bf k}$) direction rest frame spinors with positive and negative energy, respectively. Spinors at rest can be expressed in the computational basis as $\ket{00}=(1,0,0,0)^T, \dots, \ket{11}=(0,0,0,1)^T$ and represent spinors with definite (positive or negative) energy and (up or down) spin in the rest frame (for example, the first qubit index in $\ket{i,j}$ can be chosen to select positive or negative energy and the second one spin up or down). The boosted spinors can be written explicitly (see \cite{Bjorken:1964rqm}) as $|\check{\psi}_{\lambda_E,\lambda_s}\rangle  = \sqrt{m\over \omega_{\bf k}} \hat{S}_{k} |{\lambda_E,\lambda_s}\rangle$ with $(\hat{S}_{k} = {(\omega_{\bf k}+m)I-(-1)^{\lambda_E} k_i \gamma^i \gamma_0) / \sqrt{2 m (\omega_{\bf k}+m)}}$ representing the boost operator and $\lambda_E \in \{0,1\}$ and $\lambda_s \in \{0,1\}$ the eigenvalues of the energy and spin projectors, respectively. This allows us to write $\check{\rho}_k = \sum^{1}_{i,j,k,l=0} \rho_{ijkl}(k) \ket{i,j}\bra{k,l}$ and 
single out individual qubit contributions allowing for an interpretation of the functional trace, and thus of the Casimir energy, in terms of qubit degrees of freedom.
The virtual fluctuations boosted in the positive ${\bf k}$ direction are ``moving'' forward in time, while those boosted in the negative ${\bf k}$ direction are akin to positive energy modes ``moving'' backward in time - analogously to ``anti-electrons'' in the usual Dirac theory. We call these \textit{qubits} and \textit{anti-qubits}, respectively, {following a terminology used earlier in Ref~\cite{Cerf:1995sa}.}

Direct calculation of $\tr\;\check{\rho}^2(k) = \left(1-\omega_{\bf k}^2/\omega_n^2\right)/4 <1$ {(the inequality follows from the fact that $\omega_{\bf k}^2/\omega_n^2$ is real and non-negative for the present case and quantization conditions)} shows that $\check{\rho}_k$ is not a projector and, not surprisingly, it cannot be the density matrix associated to a pure state. 
Sticking with an interpretation of $\check{\rho}_k$ as the pseudo-density matrix of the virtual fluctuations, the question is whether it is separable or entangled. Performing a separability test (see Ref.~\cite{Horodecki:2009zz}) requires a number of assumptions and is not applicable to our case. It is however straightforward to prove by direct calculation that $\check{\rho}_k$ is non-separable as a single tensor product. {This can be done by considering the difference between the most general form for the tensor product between two $2\times2$ matrices and (\ref{pseudorho}), $\Delta \equiv {\mathsf A} \otimes {\mathsf B} - \check{\rho}_k$. The impossibility of writing $\check{\rho}_k$ in separable form is equivalent to showing that for any choice of ${\mathsf A}$ and ${\mathsf B}$, $\Delta$ cannot vanish identically. This can be explicitly proved by writing $\Delta$ in components $\Delta_{i,j}$ and trying to solve simultaneously all equations $\Delta_{i,j} =0$. Starting from the $\left(i,j\right) = \left(1,2\right)$ component, we find that $\Delta_{1,2} = {\mathsf A}_{11} {\mathsf B}_{12}$ (notice that $\check{\rho}_{1,2}=0$), therefore either or both ${\mathsf A}_{11}$ and ${\mathsf B}_{12}$ must vanish. Imposing either condition in all other components $\Delta_{i,j}$ shows that it is not possible to satisfy $\Delta_{i,j}=0$ identically for any choice of ${\mathsf A}$ and ${\mathsf B}$. }
This implies that $\check{\rho}_k$ cannot be written as a single direct product.

Were $\check{\rho}_k$ a canonical density matrix, this would not be a sufficient condition to prove that it is entangled. In fact, it is possible to show that in our case $\check{\rho}_k$ can be written as a convex combination of the form $\check{\rho}_k = \sum_k p_k \check{\rho}^k_1 \otimes \check{\rho}^k_2$ with $\sum_k p_k =1$, where ${\displaystyle \{\rho _{j}^{k}\}}$ for $j=1,2$ are rank-1 matrices, although not projectors in general. This suggests that $\check{\rho}_k$ is not (canonically) entangled. However, the fact that it is pseudo-hermitian (and has complex conjugate eigenvalues) makes the definition of entanglement for our pseudo-density matrix more delicate.
 
Notice also that the partially traced pseudo-density matrix $\textrm{tr}_{A}[\check{\rho}_k]$ gives $(1/2)I_{2\times 2}$, with $A$ being either the first or second qubit. 
{The partial trace operation $\textrm{tr}_{A}$ is defined in the standard form 
$\textrm{tr}_{A}[\check{\rho}_k] = \sum^{1}_{i,j,k,l=0} \rho_{ijkl}(k) \delta_{ik} \ket{j}\bra{l}$ where $A$ represents the first qubit (partial tracing over the second qubit follows a similar logic).}

The delicate point in the present case lies in the fact that the physical process at play is essentially the decay of the vacuum state into qubit-anti qubit pairs. One may expect that entanglement could occur in the presence of explicit interactions, e.g. in the presence of an electromagnetic field, but also just due to the fact that the vacuum decays into qubit-anti qubit pairs, even though each of the two sectors may be separate. A more careful study of this point is underway and will appear soon.

\textit{Casimir energy as quantum entropy.} -- The crucial step is now to express the functional determinant (\ref{fdet}) as an entropy functional. Using (\ref{pseudorhodens}) in (\ref{fdet}), we have
\bea
\Gamma^{(1)} = {-\tr\left( \check{\beta}_k \mathscr{H}_k \right)}
 -
\ln \tr\left(  e^{- \check{\beta}_k \mathscr{H}_k }\right) 
\label{eqGS}
\eea
where we have written $\ln \det$ as $\tr \ln$, and separated numerator and denominator in the logarithm {(the trace $\tr$ is over all available degrees of freedom)}. Adopting a definition of von Neumann entropy functional 
\bea
\check{\mathsf S} = - \tr \left[\check\rho_k \ln \check\rho_k\right],
\eea 
we can express $\Gamma^{(1)} = 
\check{\mathsf S}
-
\tr\left( \check{\beta}_k  \mathscr{H}_k \right) - \tr \left(\check{\beta}_k \check{\rho}_k \mathscr{H}_k \right)$. 
It is then possible to use $\tr\left( \check{\beta}_k  \mathscr{H}_k \right) = 0$ and $\tr \left(\check{\beta}_k \check{\rho}_k \mathscr{H}_k \right) = 0$\footnote{This relation can be proven by expanding in power series the logarithm 
$\tr \left(\check{\beta}_k \check{\rho}_k \mathscr{H}_k \right) =
\sum_{{\bf k},n} {\chi_k \over 2}\ln {1 + \chi_k \over 1 - \chi_k} 
= \sum_{p=0}^\infty {\mathsf{Z} \left(2p\right) \mathsf{Z}_\beta(-2p) \over 2p+1} 
=0$ and expressing the final result in terms of $\mathsf{Z} \left(q\right) = \sum_{{\bf k}} \omega_{\bf k}^{-q}$ and $\mathsf{Z}_\beta(q) = \sum_{n\in \mathbf{Z}}\omega_n^{-q}$. Then, by using the property
$\mathsf{Z}_\beta (-2p) =
({\pi/ \beta}) \sum_{n\in \mathbf{Z}}\left(2n +1\right)^{2p}\propto \zeta_H(-2p, 1/2) \to 0$, $\forall p \in \mathbf{N}$ 
($\zeta_H(u,v)$ is the Hurwitz zeta function) along with the fact that $\mathsf{Z} \left(2p\right)$ is non-singular
$\forall p \in \mathbf{N}$, gives the desired result.} to simplify (\ref{eqGS}) to
\bea
\Gamma^{(1)} = - \check{\mathsf S}.
\eea
One can note that, since the partial trace of $\check{\rho}_k$ gives the identity matrix times a factor of $1/2$, its von Neumann entropy will be $\log 2$. {Such a factor disappears after tracing out all energy and momentum modes and is added to $\check{\mathsf S}$ in order to express the Von Neumann entropy in the form of a conditional entropy:}
\bea
\check{\mathsf S} = - \tr \left[\check\rho_k \ln \check\rho_k\right] + \tr \left[\tr_A[\check\rho_k] \ln \tr_A[\check\rho_k]\right].
\eea
{The above expression is analogous to the definition of quantum conditional entropy introduced in 
quantum information \cite{Cerf:1995sa,Cerf:1999sa,Horodecki:2005vvo}.}
Specializing this result to the parallel plates setup, i.e. using (\ref{eeff}) and (\ref{parplates}) (this last relation multiplied by a factor $2$ for fermions) and normalizing the plates' area to one, we obtain a Casimir energy $\mathscr{E}_c$ 
\bea
\mathscr{E}_c = - {\pi^2\over 720 L^{3}}  = 
\lim_{\beta\to {\infty}} \beta^{-1} \check{\mathsf S}.
\eea
This result expresses the Casimir energy in terms of the quantum ``Von Neumann'' entropy $\check{\mathsf S}$ of the virtual fluctuations. In this particular setup zeta regularization has enabled us to obtain a finite result for $\check{\mathsf S}$ without the need for renormalization (due to Eq.~(\ref{eeff}) and the fact that $C_2=0$). {In curved spaces or for curved boundaries, further renormalization is expected for the Casimir energy and for the entropy.}

\textit{Conclusions} -- After reminding the known formulation of the Casimir energy as a zeta-regularized determinant, we have shown that writing the determinant in terms of a density operator allows us to introduce the concept of a pseudo-density matrix $\check{\rho}_k$, a generalization of the ordinary density matrix describing virtual quantum fluctuations. We have shown that $\check{\rho}_k$ is a two-qubit, pseudo-hermitian, matrix and that the quantum Von Neumann entropy associated to this pseudo-density matrix coincides with the quantum conditional entropy and it is precisely the Casimir energy. This approach paves the way for a comprehensive treatment of more complex setups, including curved space. Our results show that the origin of the Casimir energy can be interpreted as a measure of quantum entropy of the vacuum.

A fascinating direction to explore is the relationship between our discussions, the idea of quantum inseparability and the quantum mechanical extension of Shannon information theory, which incorporates entanglement through negative conditional entropies. This perspective involves examining quantum informational processes in which particles carry negative virtual information - specifically, quanta of negative information that can be viewed as equivalent to a qubit traveling backward in time. This concept was discussed in Refs.~\cite{Cerf:1995sa,Cerf:1999sa,Horodecki:2005vvo}. Here we observed that {the decay of the quantum vacuum state} can be interpreted as a transition from a state devoid of information to a pair of virtual states - a qubit an anti-qubit pair transporting positive and negative information, respectively. This is what our picture of the Casimir effect and the pseudo-density matrix conveys. The emergence of negative probabilities that arise from a non-canonical density matrix should be understood, in the same vein as Ref.~\cite{Feynman:1984ie}, as negative conditional probabilities that characterise intermediate states. This is clear from our entropy formula that consists of the summation of the contribution of the quantum fluctuations, not yielding a negative final probability for a physical event, instead resulting in a ``physical'' measurable outcome that is the Casimir energy.

\textit{Acknowledgements} -- We acknowledge the support of the Japanese Society for the Promotion of Science (Grant-in-Aid for Scientific Research, KAKENHI, Grant No. 21K03540) and of the Funda\c{c}\~{a}o para a Ci\^{e}ncia e a Tecnologia of Portugal (project CEECIND/02474/2018 and project EXPL/FIS-PAR/1604/2021). 
We should like to thank Prof. Karol Zyczkowski for discussions.

\end{document}